\documentclass[conference, 10pt]{IEEEtran}
\usepackage{amssymb, amsmath}
\usepackage{setspace}
\usepackage{graphicx}
\usepackage{epsfig,cite,latexsym,subfigure}
\usepackage{theorem}
\usepackage{times}
\usepackage{epsf,cite,subfigure}
\usepackage{amsmath,amssymb}
\usepackage{floatflt}
\usepackage{url}
\usepackage{times}
\usepackage[usenames]{color}

\newcommand{\E}[1]{E\left[#1\right]}

\newcommand{\set}[1]{\left[#1\right]}

\newcommand{\prob}[1]{\mbox{Pr}\left\{#1\right\}}

\newcommand{\eqnref}[1]{(\ref{eqn:#1})}

\newcommand{\eqnlabel}[1]{\label{eqn:#1}}

\newcommand     {\paren}[1]{\left(#1\right)}

\newcommand{\curlb}[1]{\left\{#1\right\}}

\newcommand{\eX}[1]{\e^{#1}}

\newcommand{\e}{e}

\title{Low Complexity Differentiating Adaptive Erasure Codes in Multimedia Wireless Broadcast}
\author{
\IEEEauthorblockN{\normalsize Silvija Kokalj-Filipovi\'c and Emina Soljanin}
\IEEEauthorblockA{\small Alcatel-Lucent, Bell Labs\\
\small\em \{silvija.kokalj-filipovic,emina.soljanin\}@alcatel-lucent.com} \and
\IEEEauthorblockN{\normalsize Predrag Spasojevi\'c}
\IEEEauthorblockA{\small Rutgers University}\\
\small\em spasojev@winlab.rutgers.edu}
\begin{document}
\maketitle
\begin{abstract}
Based on the erasure channel FEC model as defined in multimedia wireless broadcast standards, we illustrate how doping mechanisms included in the design of erasure coding and decoding may improve the scalability of the packet throughput, decrease overall latency and potentially differentiate among classes of multimedia subscribers regardless of their signal quality. We describe decoding mechanisms that allow for linear complexity and give complexity bounds when feedback is available. We show that elaborate coding schemes which include pre-coding stages are inferior to simple Ideal Soliton based rateless codes, combined with the proposed two-phase decoder. The simplicity of this scheme and the availability of tight bounds on latency given pre-allocated radio resources makes it a practical and efficient design solution.
\end{abstract}
\section{Introduction}
Multimedia Broadcast/Multicast Services (MBMS) \cite{3gppMBMS}  is a point-to-multipoint interface specification for existing and upcoming 3GPP cellular networks, designed to provide efficient delivery of broadcast and multicast multimedia content delivery, both within a cell and within the core network. It has been widely recognized that the appropriate application layer forward error correction {\em (AL-FEC)} for MBMS are adaptive coding techniques based on punctured \cite{DVBH} or rateless (Fountain) codes  \cite{r10,rq}, as their redundancy can be flexibly adapted
to different channel/network conditions.
%
%
With the proliferation of mobile video traffic, the impact of Fountain codes will be growing,
and so will the importance of decreasing both their encoding/decoding complexity and their overhead, in order to match the strict latency
constraints of streaming applications.

In this paper we explore the decoding efficiency in terms of the communication cost between the server and the client of the multimedia wireless broadcast, incurred to completely recover all data in linear decoding time.
We also address another important design challenge of wireless broadcast streaming, namely, catering to priority subscribers. Certain 3G network subscribers might not claim special bandwidth rights with their mobile providers but they may be subscribed to a multimedia streaming service with a guaranteed Quality of Service {\em (QoS)}. Hence, it is natural that these service privileges be accommodated within the application layer of the network, using the application layer FEC. The strength of Fountain codes that matters most in multimedia broadcast, and makes it scalable to a large number of clients (such as in video broadcast of popular sport events, parades, presidential debates and inaugurations) is the statistical equality of encoded symbols, not their differentiation features. We introduce a two-phase decoder that allows for differentiation while preserving broadcast-friendly features of Fountain codes.

This paper illustrates how the proposed Fountain-based adaptive FEC approach exhibits not only linear decoding time, but also a low reconstruction delay which is controlled by the client, within the framework of his QoS privileges. This mechamism leverages the peeling decoder and streamlines several existing mechanisms, including inactivation \cite{Inact} and doping \cite{JSACdopedLT}, as well as a minimal feedback. The user may opt for a peeling decoder (i.e. Belief Propagation –- BP), which is simple but the overhead is larger as we have to make sure the ripple (set of one-term equations) will never become empty, or he may choose a decoder based on Gaussian elimination (GE), which is complex but the overhead is smaller. The inactivation decoder combines BP and GE, and trades overhead for complexity. Finally, doping guarantees small overhead and linear decoding but requires minimal feedback.

One of the contributions of this work is the observation that our model of the doped peeling decoder \cite{JSACdopedLT} can be successfully applied to the peeling decoder with inactivations. Using this model, the performance bounds and their trade-offs (decoding complexity and doping communication cost) for all decoding options are clearly defined, and the user can control the trade-offs given available communication and computation resources. Most importantly, we show that complex solutions with pre-codes are not necessary, as the small-overhead linear-time decoding can be achieved by doping or inactivating a simple Ideal Soliton based code in the second phase of decoding, which is also used for differentiation. We next briefly present the usage model of Fountain codes in MBMS, which provides a motivation for our approach, and then introduce the proposed decoding mechanisms and their analysis. Section~\ref{sec:discuss} compares the cost-based performance of the Ideal Soliton code and a Fountain code whose distribution is defined by the standard \cite{r10}, using both analytical estimates and simulation results.  In section~\ref{sec:exam} we consider an example of the proposed AL-FEC use case which shows that priority users could be satisfied in a scalable fashion.
\section{Rateless Erasure Codes in Multimedia Broadcast}
\subsection{3GPP Chunked Content Distribution: Delivery and Repair}
The utilization of Fountain codes for the application layer FEC in  wireless multimedia broadcast has followed the framework proposed by the 3GPP MBMS, where the content is partitioned into source blocks (chunks usually corresponding to video frames), and each source block is further divided into $k$ source symbols of bit-length $s$. For simplicity, we will assume that each encoded symbol fits into one packet, i.e. the encoding procedure XOR-es a subset of the $k$ symbols, and encloses the resulting array of $s$ bits, termed encoded symbol, into one packet. If several encoded symbols were put into one packet, then one packet erased by the channel would affect many symbols. Although packaging optimality is a relevant problem, we abstract it here through the erasure parameter associated with the application-level transmission channel.
\begin{figure}[!t] 
\begin{center}
\includegraphics[width=3.1in]{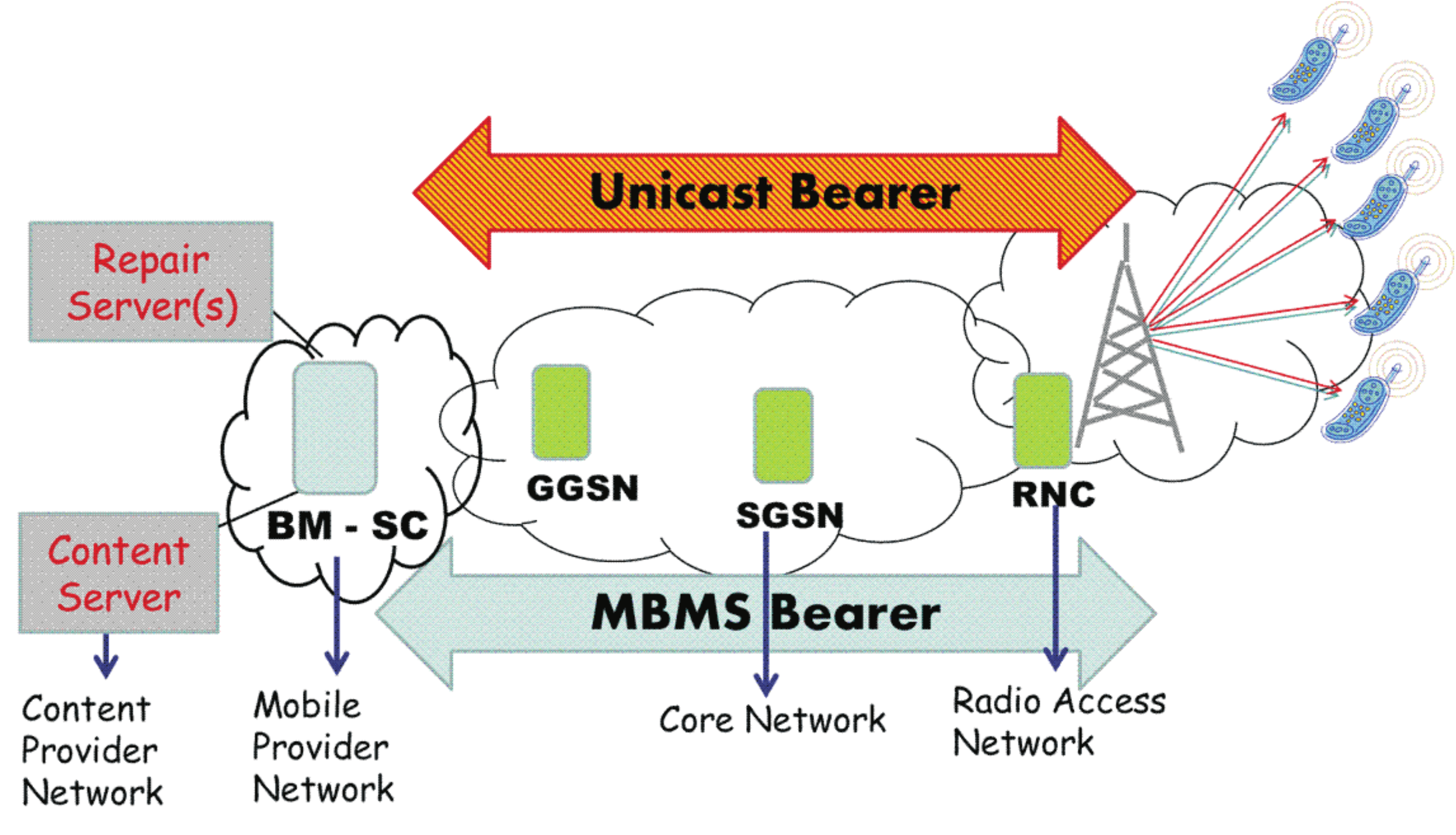}
\end{center}
\caption{3GPP MBMS Basic Architecture: the standard 3GPP entities are Radio Network Controler (RNC), Serving GPRS Support Node (SGSN) and Gateway GPRS Support Node (GGSN), while Broadcast-Multicast Service Center (BM-SC) is the MBMS specific entity. }\vspace{-6mm}
\label{fig:mbmsArch}
\end{figure}

The symbols XOR-ed in a packet represent one binary equation, where the terms (source symbol indices) are signaled in the packet header. The result of the XOR-ing, delivered as the packet load, is the value of the equation. The number and the identities (indices) of the equation terms are random, although following a given probability distribution. The 3GPP MBMS standards propose a time-limited {\em delivery phase} in which to first broadcast $k$ packets, each carrying one source symbol of the block i.e. a distinct one-term (singleton) equation, and then a number of parity-check symbols (higher-degree equations). The MBMS framework is seamlessly incorporated into existing 3GPP architecture, with the exception of some services designed for evolved 3GPP networks only (i.e. 3G Long Term Evolution, or 3G LTE), such as broadcasting in MBMS single frequency networks (MBSFN). We do not consider such services, following an assumption that the entire system will be evolving further, and, hence, our goal is to use the MBMS basic framework as an abstract platform only to demonstrate usefulness of the proposed approach. For a complex analysis of the standard-based MBMS AL-FEC readers are referred to \cite{StandardAL-FEC}.

The basic MBMS includes dedicated channel resources (MBMS radio bearers) used to broadcast multimedia content to multiple wireless receivers. Figure~\ref{fig:mbmsArch} illustrates the basic MBMS architecture in which $BM-SC$ denotes {\em Broadcast/Multicast Service Center,} a logical entity that controls seamless broadcast from the content servers by coordinating between the 3GPP radio resource allocation controllers, and the streaming data users. Apart from the MBMS radio bearers, the radio resources include unicast (interactive) channels from the so called repair servers to multimedia wireless users.

The scenario in which during the delivery phase each wireless broadcast client collects a set of encoded symbols resulting in a solvable system of equations is not very likely.
Hence, once the delivery (broadcast) session expires, a unicast-based file repair mechanism is available in the post-delivery phase. Despite the expected uniform distribution of repair sessions,
the fact that the server potentially serves many requests may cause a communication bottleneck. For that reason, we believe that the repair mechanism must be accounted for in the design of the coding scheme, assuming a certain deterministic order in serving repair requests to allow for a good QoS control, and to mitigate the fact that the high-priority users may be handicapped by the low SNR. Let us denote by $\Delta_f$ the feedback delay, which
quantifies the overhead resources used for communication to
and from the repair server. Specifically, to communicate with the repair server, each user has to establish a context switch facilitated by the BM-SC in both physical layer, and the upper protocol layers, which includes allocating a different radio bearer (for unicast), and coordination among many network management instances. In addition, $\Delta_f$ includes the service waiting time with the repair server.  Consequently, note that the priority-based serving order will cause the expected value of $\Delta_f$ to vary according to the privileges of a specific user. As for the context switching delay, we here assume that it is a fairly deterministic but significant part of $\Delta_f$. To assess the implications of the repair system latency, we next describe the trade-offs in the communication overhead.
\subsection{Successful Decoding: Over-designing Communication Overhead vs Allowing Repairs}
For a high-quality multimedia delivery, the decoding failure probability must be constrained to zero. Hence, we seek to quantify the cost in terms of the communication overhead for an application-level FEC that does not allow for any undecoded symbols.  A related performance measure, prominent in the analysis of practical Fountain codes \cite{RaptorMonograph}, is the  overhead-failure curve, describing the failure probability $f(o)$ as a function of the overhead $o.$ Here, $o=n-k,$ where $n$ is the number of collected encoded symbols. Typically, $f(o)$ is a quickly decreasing function. In case of random Fountain codes, where a random number of uniformly selected source symbols is combined in each encoded symbol, the failure probability is easy to calculate as the probability that this random set of equations is not of full rank, and can be bounded by $2^{-o}$  \cite{Kolchin-RG}. However, random Fountain codes do not satisfy the multimedia latency requirements as the decoding complexity is high. The linear decoding time may be achieved by optimizing probability distribution of the number of terms the equations have. This number of terms is often called the output symbol degree. Luby {\em (LT)}  codes \cite{luby} have become popular thanks to such an optimized distribution, i.e. the {\em Robust Soliton (RS)}, which promises  linear decoding time. The RS is a design that grew out of the {\em Ideal Soliton (IS)} distribution \eqnref{isol}, which was the ideally linear distribution (in terms of average decoding time) obtained analytically. To compensate for the variance of the empirical distribution of sampled symbol degrees, which may cause the linear decoder to stall, the RS design moves some probability mass from the higher degrees to degree one. As a result, the empirical decoding time of LT codes is close to linear. However, to achieve acceptable failure probability LT code design required a sizable overhead. This motivated the design of another popular rateless code, dubbed Raptor \cite{Raptor}, which combines a pre-code stage with LT encoding to generate the output symbols decodable with constant overhead. This more complex design is difficult to rigorously analyze, and, instead, some heuristics are used to optimize the performance \cite{RaptorMonograph}.

The overhead may be decreased if we allow for a repair procedure to identify and fetch the missing symbols, given strictly limited overhead in the upfront collected output symbols. The symbols missing to reconstruct the entire source block from the collected equations can be identified through an attempted decoding procedure. The decoding can be iterative, i.e. a message-passing erasure (peeling) decoder, or it can rely on classical algorithms for solving linear system of equations, such as {\em Gaussian elimination (GE)}. A system of equations solvable through GE may not be solvable by iterative decoding. Even though the GE-based decoder is optimal, its complexity may be prohibitive. Rateless codes should be designed so that all input symbols can be recovered with high probability using an iterative decoder on a set of equations (collected coded symbols) slightly larger than $k$. We here consider only the iterative decoder, as multimedia latency constraints dictate linear decoding time. Given a {\em peeling decoder (PD)}, the repair symbols can be determined in a sequential manner \cite{JSACdopedLT}. Here, if the decoder stalls, an assisting procedure identifies a symbol capable of repairing (doping) the decoder, and immediately requests it from the server.
\subsection{Communication Overhead of the Repair Process}
We here specify repair communication overhead in terms of bit-delay equivalents.  To distinctly specify the identified source symbol to the repair server, we need $\log{k}$ bits. Adding the bits that the server uses to transfer one vector symbol from the field $\mathbb{F}_q$ (of cardinality $q$), this makes $\log{k}+s\log{q}$ bits of per-symbol repair cost. As a source block is most frequently equivalent to a video frame (of size $b=1MB$), and we assume that $k\geq 1000,$ , and $q=2,$ note that $s=b/k\leq 1MB/128B \leq=8000$ bits. The sequential repair (i.e. every time the peeling decoder stalls) incurs the total per-symbol cost of $\log{k}+s\log{q}+\Delta_f,$ where $\Delta_f$ is the bit-equivalent feedback roundtrip delay. In this paper, we propose sequential identification of repair symbols, while avoiding sequential repair (doping). The doping symbols will be considered free variables to be revealed at the end. This postponed-repair design, akin to \cite{Inact}, allows for complete linear decoding safe for a set of symbols that will be either requested from the repair server at the end of the procedure, or solved by Gaussian Elimination, or the combination of the two. Our stochastic model of the decoding procedure puts a tight bound on the number of symbols that must be repaired, and demonstrates that a simple encoding procedure based on Ideal-Soliton distribution of equation degrees yields a diminishingly small repair overhead.

Let us denote the percentage of the undecoded symbols by $M,$ which is a random variable. The {\em per symbol} cost of this postponed repair would then amount to $\log{k}+s\log{q}+\frac{\Delta_f}{Mk}.$ This lowers the cost with respect to plain doping \cite{JSACdopedLT}, while still maintaining the linear decoding time. We show, both analytically and through simulations, that a simple Fountain code, with well designed linear decoder, results in $M\approx 1\%.$  Hence, both the per-symbol and the total communication overhead can be made relatively small, since usually $k\geq 1000$.
\section{Design Preliminaries}
In
\cite{SanghaviRtless}, the author shows that
the recoverable fraction of input symbols depends on the output degree
distribution of the code.
The results in \cite{SanghaviRtless} are of interest for real-time systems using
rateless codes, including multimedia wireless broadcast. Apart from emphasizing the importance of the output degree distribution, they imply that if the erasure rate is above a certain value, given the limited duration of the session, the collected system of equations will not be sufficient for iterative decoding under any distribution. This motivates the extensions to the iterative decoder, presented in Section~\ref{sec:enhanced}, and assisted by doping.

In order to establish a tight bound on the communication cost, we focus in this paper on pure LT codes. Moreover, we consider LT codes based on the IS, as it allows for a straight-forward analysis of the occasional assistance to the decoding process when it gets stalled \cite{JSACdopedLT}. The availability of this assistance obviates the need for overhead-failure analysis, as we are allowed to get additional symbols on demand, i.e. to keep doping a minimum set of equations until it reaches full rank.
In addition, we consider the LT codes used in the standardized Raptor designs, but not in their systematic form.
The assumed existence of clients that cannot decode at all was the motivation behind the choice of the systematic structure of the MBMS standardized Fountain (Raptor) codes \cite{r10,rq}, where some of the encoded symbols are equivalent to the source symbols (singleton equations), and hence, the decoding is trivial at the expense of complex encoding. We argue that multimedia clients must have decoding capabilities, or otherwise expect only the best-effort service. Besides, the systematic structure compromises the concept of ratelessness in terms of the statistical equality of encoded symbols. Let us point out that the systematic implementation is compelling only for erasure-free channels, as otherwise, if the receiver can handle only the systematic symbols, the single eligible rateless code is the repetition code, which is inefficient.

The standardized Raptor proposes two mechanisms that combine iterative decoding with Gaussian elimination \cite{Inact} so that the complexity remains linear, while the collected set is more likely to be sufficient for decoding with a slight but acceptable complexity increase.
When combined with these {\em inactivation mechanisms}, iterative decoder is allowed to continue until all of $D=Mk$ repair symbols are identified, and then send a single doping request at the end. Upon receiving the requested symbols, the decoder can completely recover the source block in linear time by back-substituting the ``doped'' symbols. The delay in decoding stems only from the repair latency at the end.

We next present the analytical model of the peeling decoder assisted by doping, before describing our implementation of the inactivation mechanisms based on the two flavors of LT code (the IS, and the standardized Raptor distribution). With Raptor implementation, we omit the pre-code stage, given that the inactivation mechanisms in the peeling decoder (PD) play the role of a pre-code in decreasing the probability of failure. Besides, this decreases the complexity, which is of paramount importance for mobile applications, and simplifies the analysis. Our analytical results and simulations justify this approach as the repair cost is lower when compared with the plots presented for standardized Raptor codes (with pre-codes) in \cite{RaptorMonograph} (Figure 3.4, pg. 270).
\section{Solution: Enhanced Peeling Decoders}\label{sec:enhanced}
One of the contributions of this work is the observation that our model of the doped peeling decoder \cite{JSACdopedLT} can be successfully applied to the peeling decoder with inactivations. We next present the adapted model.
\subsection{Model of the Basic Peeling Decoder}
\begin{figure}[!t] 
\begin{center}
\vspace{-8mm}
\includegraphics[width=3.5in]{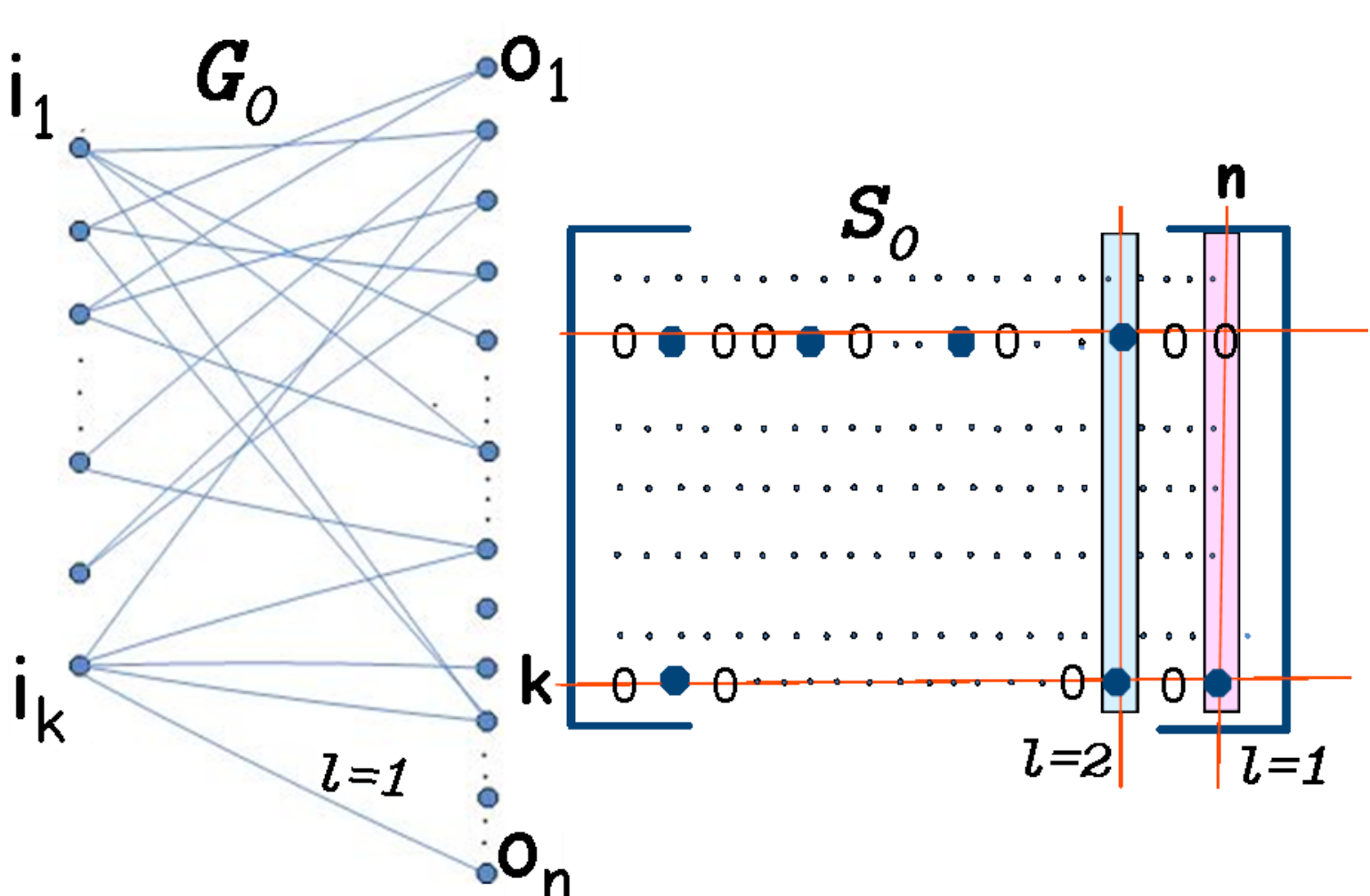}
\vspace{-4mm}
\end{center}
\caption{Peeling decoder manipulates the incidence matrix corresponding to the code graph. The first two decoding steps are shown ($\ell=1,\ \ell=2)$ which erase two rows in the initial matrix $\mathbf{S_0},$ and two edges in the graph $\mathbf{G_0}$.}\vspace{-6mm}
\label{fig:decstart}
\end{figure}

Let us have a set of $k_s$ code symbols that are linear combinations of $k$ unique input symbols, indexed by the set $\curlb{1,\cdots,k}$. Let the degrees of linear combinations be random numbers that follow  distribution $\omega(d)$ with support $d\in\curlb{1,\cdots,k}$. We equivalently use $\omega(d)$ and its generating
polynomial $\Omega(x)=\sum^{k}_{d=1}{\Omega_dx^d},$ where
$\Omega_d=\omega(d).$  Let us denote the graph describing the peeling
 decoding process at time $t$ by $\mathbf{G_t}$ (see Figure~\ref{fig:decstart} depicting the graph at $t=0,$ for $k_s=n$).
We start with a decoding matrix $\mathbf{S_0}=\set{s_{ij}}_{k\times k_s},$ where code symbols are described using columns, so that $s_{ij}=1$ iff the $j$th code symbol contains the $i$th input symbol. Number of ones in the column
corresponds to the degree of the associated code symbol. Input
symbols covered by the code symbols with degree one constitute the
ripple. In the first step of the decoding process, one input symbol
in the ripple is processed by being removed from all neighboring code symbols in the associated graph $\mathbf{G_0}$.
If the index of the input symbol is $m$, this effectively removes
the $m$th row of the matrix, thus creating the new decoding matrix
$\mathbf{S_1}=\set{s_{ij}}_{(k-1)\times k_s}.$ We refer to the code
symbols modified by the removal of the processed input symbol as
{\em output symbols.} Output symbols of degree one may cover additional input
symbols and thus modify the ripple. Hence, the output degree distribution changes to $\Omega_1(x)$.

At each subsequent
step of the decoding process one input symbol in the ripple is
processed by being removed from all neighboring output symbols and all such output symbols that subsequently have exactly one remaining neighbor are released to cover that
neighbor. Consequently, the support of the output symbol degrees
after $\ell$ input symbols have been processed is
$d\in\curlb{1,\cdots,k-\ell},$ and the resulting output degree
distribution is denoted by $\Omega_{\ell}(x)$.
Since the encoded symbols are
constructed by independently combining random input symbols, we can assume that the input symbols covered by the degree-one symbols are selected uniformly at random from the
set of undecoded symbols. Hence, we model the $\ell$th step of the decoding process by selecting a row
uniformly at random from the set of $(k-\ell)$ rows in the current
decoding matrix $S_{\ell} = \left[s_{ij}\right]_{(k-\ell)×k_s}$, and removing it from the
matrix. After $\ell$ rounds or, equivalently, when there are $k - \ell$ rows in the decoding matrix, the number of non-zero coefficients in a column is denoted by $A_{k-\ell}$. The probability that the column is of degree $d,$ when its length is $k-\ell-1,\ \ell\in\curlb{1,\cdots,k-3}$, is described iteratively
\begin{eqnarray}\eqnlabel{probd}
\nonumber P\paren{A_{k-\ell-1}=d}&=&P\paren{A_{k-\ell}=d}\paren{1-\frac{d}{k-\ell}}\\
&+&P\paren{A_{k-\ell}=d+1}\frac{d+1}{k-\ell}
\end{eqnarray}
for $2\leq d < k-\ell,$ and $P\paren{A_{k-\ell-1}=k-\ell}=0.$

Let the starting distribution of the column degrees (for the decoding matrix $\mathbf{S_0}=\set{s_{ij}}_{k\times k_s}$) be denoted by $\omega_0(d).$
By construction, for $l=0,$ $P\paren{A_{k}=d}=\omega_0(d),$ which, together with \eqnref{probd}, completely defines the dynamics of the decoding process.

Let $k_s =
k (1 + \delta),$ where $\delta$ is a small positive value. At time $\ell$ the total
number of decoded and doped symbols is $\ell,$ and the number
of output symbols is $n = k_s - \ell = \lambda^{\delta}_{\ell}
(k - l).$
Here,
$\lambda^{\delta}_{\ell} = 1 +\frac{k}{k-\ell}\delta$ is an increasing function of $\ell$. The unreleased output symbol degree distribution polynomial at time
$\ell$ is $\Omega_{\ell}(x)=\sum{\Omega_{d,\ell}x^d},$ where
$d=2,\cdots,k-\ell,$ and $\Omega_{d,\ell}=\omega_{\ell}(d).$
Each decoding iteration processes a random symbol of
degree-one from the ripple. Released output symbols are its
coded symbol neighbors whose output degree is two. Releasing output symbols by processing a ripple symbol corresponds to performing, in average, $n_2 = n\Omega_{2,\ell}$ independent Bernoulli
experiments with probability of success $p_2 = 2/(k - \ell).$
Hence, the number of released symbols (or the ripple increment) at any decoding
step $\ell$ is modeled by a discrete random variable $\Delta
^{(\delta)}_{\ell}$ with Binomial distribution $\mathcal{B} (n\Omega_{2,\ell}, 2/(k - \ell)),$ which for large
n can be approximated with a (truncated) Poisson distribution. In \cite{JSACdopedLT} we model the ripple process as a random walk, i.e. a partial sum of shifted Poisson random variables, and analyze the stopping time of this process. Readers interested in detail analysis are referred to Appendix of \cite{JSACdopedLT}.
\subsubsection{The Ideal Soliton Advantage}
Let the starting distribution of the column degrees (for the decoding matrix $\mathbf{S_0}=\set{s_{ij}}_{k\times k_s}$) be Ideal Soliton, denoted by $\rho(d),$
\begin{equation}\eqnlabel{isol}
\rho(d)=\frac{1}{d(d-1)} \mbox{\quad for $d=2,\cdots,k$},
\end{equation}
and $\rho(1)=\frac{1}{k}.$
After rearanging and canceling appropriate terms in \eqnref{probd}, we obtain, for $d\geq 2,$
\begin{eqnarray}\eqnlabel{genersimple}
P\paren{A_{k-l}=d}&=&
\begin{cases}
\frac{k-l}{k}\rho(d) & d=2,\cdots,k-l,\\
0 & d > k-\ell.
\end{cases}
\end{eqnarray}
We assume that $k_s\approx k$ as, by design, we desire to have the set of upfront delivered symbols $k_s$ as small as the set of source symbols. The probability of degree-$d$ symbols among  $k_s-\ell$ output symbols can be approximated with $$\frac{P\paren{A_{k-\ell}=d}k_s}{k_s-\ell}\approx\frac{P\paren{A_{k-\ell}=d}k}{k-\ell}.$$
Hence, the probability distribution
$\omega_{\ell}(d)$ of the unreleased output node degrees at any
time $\ell$ remains the Ideal Soliton
\begin{equation}\eqnlabel{genersimpleU}
\omega_{\ell}(d)=\frac{k}{k-\ell}P\paren{A_{k-\ell}=d}=
\rho(d) \mbox{\quad for $d=2,\cdots,k-\ell$} .
\end{equation}
This stationary character of the IS based decoding induces the IID (Independent Identically Distributed) nature of the ripple increment, as, according to \eqnref{genersimpleU}, the fraction of degree-two output
symbols for the IS based Fountain code is expected to be
$n_2/n \approx \Omega_{2,\ell}=\rho(2)=1/2,$ for any decoding
iteration $\ell.$ Hence,
\begin{equation}\eqnlabel{releasedIdealdelta}
\eta(r)=\prob{\Delta^{(\delta)}_{\ell}=r}=\frac{\paren{\lambda^{(\delta)}_{\ell}}^r\eX{-\lambda^{(\delta)}_{\ell}}}{r!},\,\,\,  r=0,\cdots,\frac{n}{2}
\end{equation}
or, equivalently,
$\Delta^{(\delta=0)}_{\ell} \sim \wp\paren{1},$ where $\wp\paren{\cdot}$ denotes Poisson distribution.
With ripple increment of the IS decoding being an IID Poisson of unit mean, the analysis of the stopping time  as their partial sum is straightforward, and results in a tight bound of doping frequency.

Otherwise, the analytical models for
ripple evolution, characterizing the decoding of LT codes with generic distribution $\Omega_0(d),$ are extremely complex. The distribution of the $c$ output symbols in the {\em cloud} (i.e. the symbols of degree larger than one) can only be characterized through the joint non-stationary distribution of the ripple of cardinality $r,$ and cloud of size $c$, $\Omega_{\ell}(x,y)=\sum_{c\geq 0, r\geq 1}{\Omega_{c,r,\ell}x^cy^{r-1}},$ at any step $\ell$ \cite{karpLTanalysis, maneva}. As a result, the stopping time of the ripple is hardly tractable.
The stopping time corresponds to the event of empty ripple, which would mean the failure of the decoding process, if it weren't for the possibility of doping.
\subsection{Model of Doping and Inactivation}
With doping, we define
$\curlb{T_i}$ as a sequence of
stopping-time random variables where  index  $i$ identifies a doping
round. $Y_i=T_i-T_{i-1},i>1$ is the stopping time interval, equivalent to the number of decoded symbols between dopings or {\em interdoping
yield.}
The interdoping yield is evaluated using the following recursive expression
\begin{eqnarray} \eqnlabel{RecursnonIID}
\prob{Y_i=0}&=&\prob{Y_i=1}=0\\
\nonumber \prob{Y_i=t+1}&=&\eta(0)R^{\eta}\paren{t}\mbox{\quad $1\leq t<k$},\\
\nonumber R^{\eta}\paren{t}=\aleph^{(t)}(t-1)&-&\sum^{t-1}_{i=1}{\prob{Y_i=t-i}\aleph^{(i)}(1+i)}
\end{eqnarray} 
Here, $\eta(0)$ is Poisson pdf of intensity
$\lambda^{(\delta)}_{T_{i-1}}=1+\delta\frac{k}{k-T_{i-1}},$ evaluated at $0,$ and
$\aleph^{(s)}(d)$ is the $s$-tupple convolution of $\eta(\cdot)$
evaluated at $d,$ resulting in a Poisson pdf of intensity
$s\lambda^{(\delta)}_{T_{i-1}}$ evaluated at $d.$
In special case when $\delta=0,$ further simplifying assumptions
lead to the approximation that all interdoping yields are described
by a single random variable $Y$ whose pdf is given by the following
recursive expression, based on \eqnref{RecursnonIID},
\begin{eqnarray} \eqnlabel{Recurs}
&&\prob{Y=t+1}=\\
\nonumber &&\eta(0)\paren{\wp^{(t)}(t-1)-\sum^{t-1}_{i=1}{\prob{t-i}\wp^{(i)}(1+i)}},
\end{eqnarray}
where $\wp^{(s)}(d)$ denotes Poisson distribution of intensity $s,$
evaluated at $d,$ and $t\in[0, \,\, k-1].$
Now, the expected value of
the
interdoping yield $Y$ is 
\begin{eqnarray}\eqnlabel{parsum}
\E{Y}\approx\sum_{t=1}^k{t\prob{Y=t}} - \left(1-
\sum_{t=1}^k{\prob{Y=t}}\right) k.
\end{eqnarray}
Finally, the doping process is a renewal process (ignoring the final stages when $\ell \approx k$), and thus,
the Wald Equality
\cite{Gallager-DSP} implies that the expected number of dopings, i.e the additional singletons the decoder needs to obtain to complete the peeling process, is
\begin{eqnarray}\eqnlabel{edop}
\E{D}= k/\E{Y}.
\end{eqnarray}
 We may use several techniques to select these singletons. The best and most tractable results are obtained with {\em degree-two doping}, choosing the symbols present in the remaining degree-two equations, which makes decoding and doping steps indistinguishable in terms of ripple increments. Evaluation of $\E{D}$ for relevant values of $k$  (i.e. $\geq 1000$) shows that dopings are on the order of $1\%$ (see Fig.~\ref{fig:syman}). A recent contribution, based on our model of the ripple process, analyzes several other doping mechanisms, and their usage for wireless broadcast \cite{DopedLTWless}.

The concept of inactivation in the decoding of rateless codes has been introduced in \cite{Inact}. We distinguish {\em dynamic inactivation (DI)} from {\em permanent inactivation (PI)}. We observe that the dynamic inactivation has the stochastic properties of the presented random walk model for doping, as an instance of DI occurs under the same conditions as the doping, i.e. when the decoding process stalls.
\subsubsection{Dynamic Inactivation}
The basic idea behind DI is to designate a source symbol
in the decoding matrix as decoded but of an unknown value whenever an empty ripple occurs. Assigning the unknown value is equivalent to introducing a free variable in the solution of the system of equations. Let us utilize our matrix $S_{\ell}$ to explain how the DI can be implemented to restart the PD, for the first time, at the decoding step $\ell.$ One of the ways to mark a source symbol $x_q$ as "decoded" for the remainder of
the peeling process is to add an extra (empty) row to $S_{\ell}$ corresponding to a free variable $z_1,$ and then expand this modified $S^{(1)}_{\ell}$ with another column containing ones only at the positions $q$ and $k-\ell+1.$ The codeword is also extended with a zero symbol at the position corresponding to the added column, which models the equation $x_q+z_1=0$ (in GF2). To restart the PD for the $v^{th}$ time at the step $p$, we extend $S^{(v)}_p$ with additional column. The symbol $x_q$  marked
as "decoded" is chosen in such a way that a new ripple symbol is released allowing the PD
to continue (any of the two circled symbols of column $j$ in Figure~\ref{fig:dynact}). Now, the symbol is not being released in the way it happens with decoding or doping. That is, in the matrix equivalent, the $q^{th}$ row is not erased entirely. Instead, all unit coefficients in this row (corresponding to coded symbols where the $q^{th}$ source symbol appears) are replaced by zeros, and ones are written in the respective columns at the added row (see the vertical arrows depicting this modification in Figure~\ref{fig:dynact}). Hence, from the moment of first inactivation, the free variables are percolating the columns, making every consequent release dependent on the value of free variables. The completion of this modified peeling process results in a decoding matrix with the block structure presented in Figure~\ref{fig:dynact}. We permute columns of the matrix to have ones in the upper-left block appear on the diagonal, making it an identity matrix describing the source symbols, while the upper right corner is an all zero matrix (as in the upper submatrix of Figure~\ref{fig:perminactF}). The bottom submatrix describes the free variables, reflecting the dependence of the solution upon these unknown values. The values of the dynamically inactive symbols can be determined by Gaussian elimination. Assuming that the number of DIs is small (by eqn~\eqnref{parsum}, as shown in \cite{JSACdopedLT,DopedArq}), and the matrix is of full rank, the superquadratic complexity term of this last stage of the decoding does not prevail, and the overall decoding complexity is linear. If the matrix is not of full rank, we dope the symbols that have been dynamically inactivated.
\subsubsection{Permanent Inactivation}
One of the main novelties introduced with the Raptor-Q variant of standardized Raptor codes \cite{RaptorMonograph} is the use of
permanent inactivation (see Figures~\ref{fig:perminactF}~and~\ref{fig:perdynact}). We here describe permanent inactivation (PI) and analyze the impact of this technique on the decoding linearity and the communication overhead when combined with dynamic inactivation (DI).
With PI, the degree distribution of the initial matrix is changed. For any column (equation), we select $d$ random symbols from the first $k-p$ rows (source symbols), where $d$ is sampled from the probability distribution $\Omega(d),$ with support $\curlb{1,\cdots ,k-p}$. The rest of the rows contribute to the overall degree of the column according to a uniform distribution $U^{(p)}$ over the range $\curlb{1,\cdots ,p}.$ The right-hand side of Figure~\ref{fig:perminactF} depicts the sampling process, while the upper matrix in Figure~\ref{fig:perdynact} shows the initial matrix. The decoding process is illustrated in the lower matrix of Figure~\ref{fig:perdynact}, while the left hand side of Figure~\ref{fig:perminactF} shows the end result of decoding with PI and DI (after permutations, and before doping). Note that the structure of the final matrix does not differ from the case without PI if PD is applied in {\em conditional} mode, explained in subsection~\ref{subsubsec:modes}. The only notable difference is that the identity matrix is of size $k-p \times k-p,$ and, hence,  the bottom part is thicker.

We take $p$ to be on the order of $\sqrt(k),$ to maintain linear decoding complexity, while improving the matrix rank. It is known \cite{BerlekampECC80} that random matrices have a better rank profile than sparse matrices (such as LT generator matrix). The probability of full rank of a pure binary random matrix of size $p\times p+m,\ m\geq 0,$ and sufficiently large $p,$ is
$Q_m=\prod^\infty_{i=m+1}\paren{1-\frac{1}{2^{i-1}}}.$ It turns out that sufficiently large $p$ is on the order of 10. Otherwise, we calculate the probability of full row rank $p$ according to
\begin{eqnarray}\eqnlabel{frankprob}
Q_m=\prod^{p-1}_{i=0}\paren{1-\frac{1}{2^{p+m-i}}}.
\end{eqnarray}

Our simulations show that sampling the degrees by distribution $\frac{k-p}{k}\Omega(.)+\frac{p}{k}U^p(.)$ will result in an improved rank profile of the upfront-delivered set of equations (see the green curves in the close-up in Figure~\ref{fig:dopPer}, Section~\ref{sec:discuss}, depicting the decreased number of uncovered symbols, one of the main manifestations of rank deficiency).
\subsubsection{Decoding Modes}\label{subsubsec:modes}
If the initial matrix has a form presented in the upper part of Figure~\ref{fig:perdynact}, i.e. $p>0,$ we propose to apply the peeling decoder only to the rows that are not permanently deactivated, as if the symbols associated with permanent rows are  given to us as side information. We refer to such decoding as conditional, implying that it is conditional on the knowledge of the last $p$ rows. Let us refer to the first $k-p$ symbols of a column as the upper subcolumn, and the last $p$ symbols as the lower subcolumn. Similarly to decoding with DI only, releasing a degree-one upper subcolumn results in propagating the non-zero coefficients from the lower subcolumn to all the columns containing the released source symbol.
The first $p$ rows of the submatrix $D$ (Fig.~\ref{fig:perminactF}), created after the permutation of columns, define a submatrix $D_p$ of size $p\times w,$ where $w=k_s-(k-p-u)+i,$ $k_s$ is the number of upfront delivered symbols, and $i$ is the number of DIs.

It is clear that $D_p$ is a thick random matrix, i.e. $w>p$ even if $k_s=k,$ due to DIs. Our simulations also confirm that $D_p$ is of full rank with high probability. Hence, the permanently inactivated symbols can be solved by GE of small complexity, given that $p$ is on the order of $\sqrt{k}.$ This justifies the conditional decoding method, i.e. the fact that we consider permanently deactivated rows as known side information. $D_i,$ the lower part of the submatrix $D,$ consisting of $i$ rows, hence, of dimensions $i\times w,$ is created by dynamic inactivations, regardless of the existence of permanently inactivated equations. Its rank is discussed in Subsection~\ref{subsec:rankdef}.

In the unconditional mode of decoding we run PD over all matrix rows, and hence, the decoding process is plain PD as long as there are no dynamic inactivations. If a dynamic inactivation occurs, the procedure is the same for both modes: a row is added at the bottom of the matrix, and then a column is appended with unit coefficients in the inactivated row and in the added one. According to our simulations, the number of inactivations, and hence the overhead, is much larger for the unconditional mode. This is expected, as the degree distribution of columns with permanent inactivations deviates from Ideal Soliton.%
%
 \begin{figure}[!t] 
\begin{center}
\vspace{-4mm}
\hspace{-14mm}\includegraphics[width=4.4in]{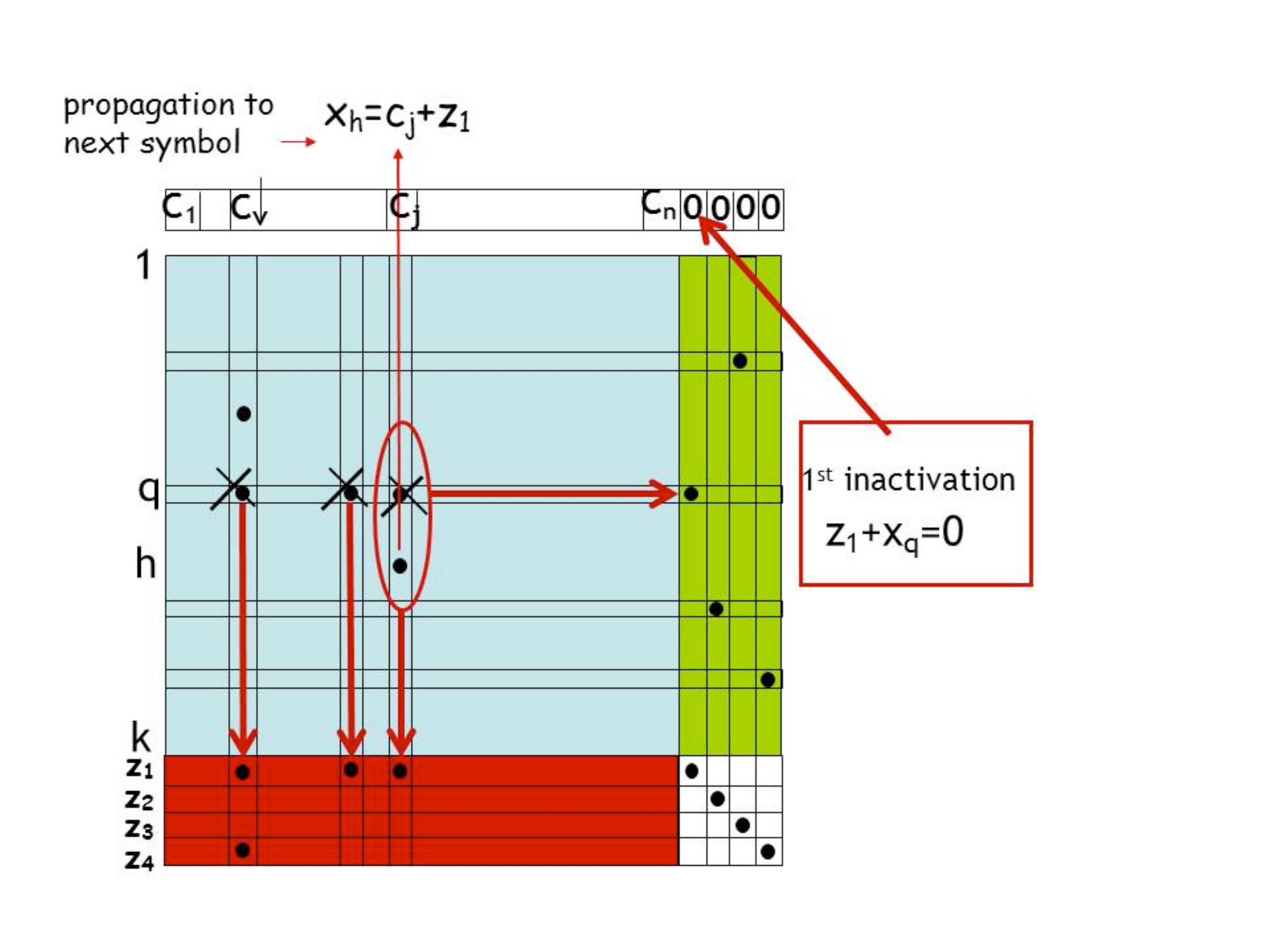}
\vspace{-6mm}
\end{center}
\caption{The initial matrix $\mathbf{S_o},$ as changed due to addition of rows matching free variables and columns matching dynamic inactivations.}\vspace{-4mm}
\label{fig:dynact}
\end{figure}
 \begin{figure}[!t] 
\begin{center}
\vspace{3mm}
\hspace{-10mm}\includegraphics[width=3.2in]{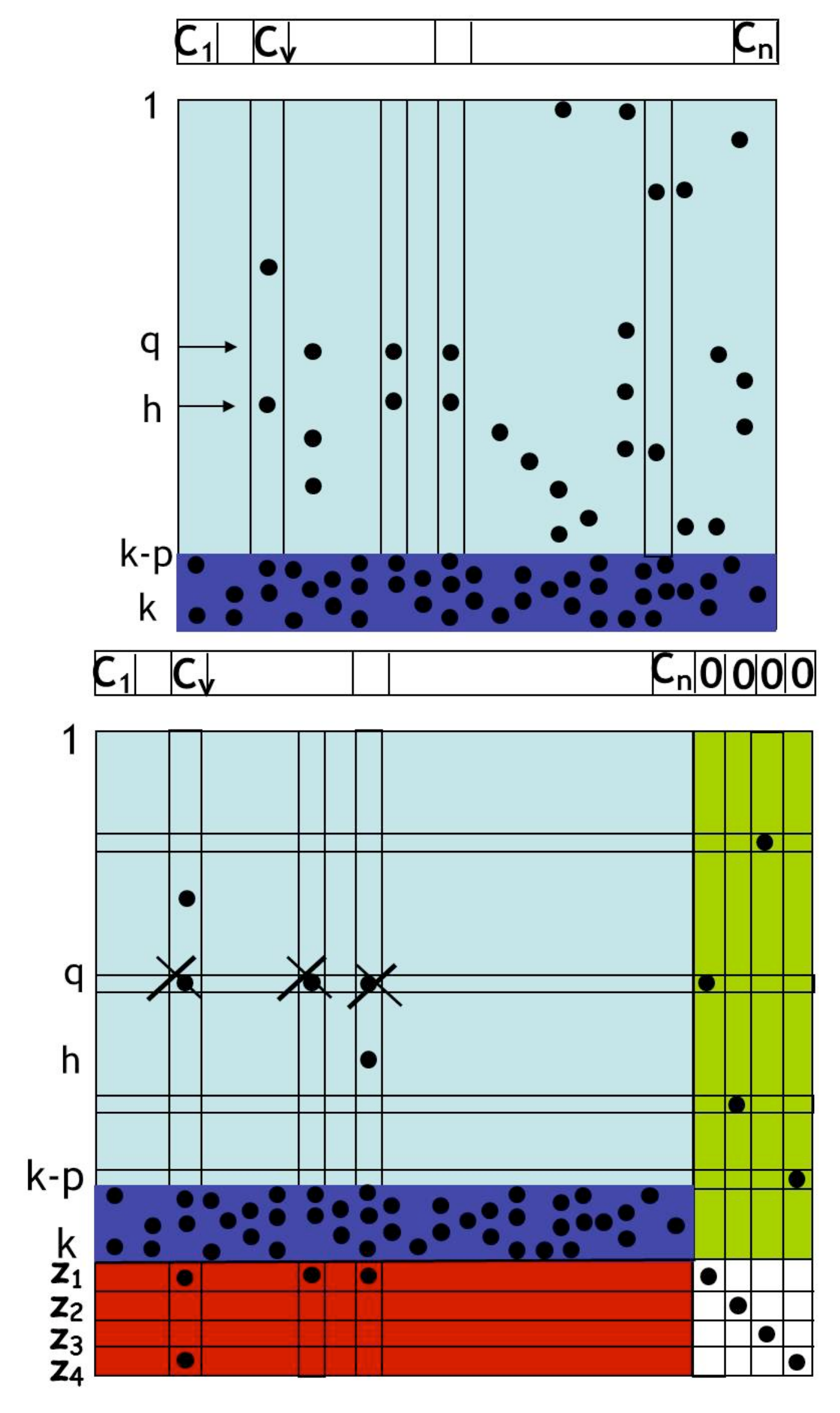}
\vspace{-4mm}
\end{center}
\caption{Changes induced in the matrix structure by the peeling decoder with dynamic inactivations over permanent inactivations.}
\vspace{-3mm}
\label{fig:perdynact}
\end{figure}
\begin{figure}[t] 
\begin{center}
\begin{tabular}{c c}
\begin{tabular}{c}
\begin{picture}(220.0,100.0)
\put(-25,90){\makebox(-2,-2)[bl]{\scriptsize{k-p}}}
\put(-20,240){\makebox(-2,-2)[bl]{\scriptsize{1}}}
\put(0,100){\framebox(220.0,140.0)[bl]{}}
\put(140,100){\framebox(80.0,140.0)[bl]{}}
\put(20,220){\circle*{4.500000}}
\put(40,200){\circle*{4.500000}}
\put(60,180){\circle*{4.500000}}
\put(65,150){\makebox(20,10)[bl]{I}}
\put(80,160){\circle*{4.500000}}
\put(100,140){\circle*{4.500000}}
\put(120,120){\circle*{4.500000}}
\put(90,150){\makebox(20,10)[bl]{\LARGE{G}}}
\put(180,150){\makebox(20,10)[bl]{O}}
\end{picture}\\
\begin{picture}(220.0,60.0)
\put(5,85){\makebox(-2,-2)[bl]{\scriptsize{1}}}
\put(220,85){\makebox(-2,-2)[bl]{\scriptsize{n}}}
\put(-35,100){\makebox(-2,-2)[bl]{\scriptsize{k(+i)}}}
\put(0,100){\framebox(220.0,60.0)[bl]{}}
\put(140,100){\framebox(80.0,60.0)[bl]{}}
\put(90,120){\makebox(20,10)[bl]{\LARGE{P}}}
\put(180,120){\makebox(20,10)[bl]{D}}
\end{picture}
\end{tabular}&
\hspace{-0.23in}
\includegraphics{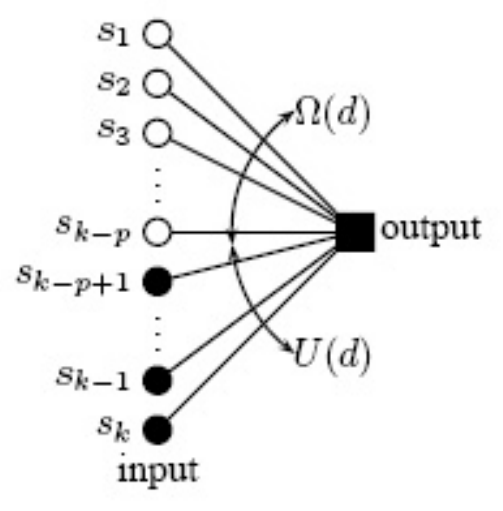}
\end{tabular}
\vspace{-0.25in}
\caption{
Degree of the terms belonging to the first $k-p$ rows (source symbols) is sampled from the prob. dist. $\Omega(d),$ while the rest of the rows contribute to the degree of the column according to a uniform dist. of range $\curlb{1,\cdots ,p}$} \label{fig:perminactF}\vspace{-2mm}
\end{center}
\end{figure}
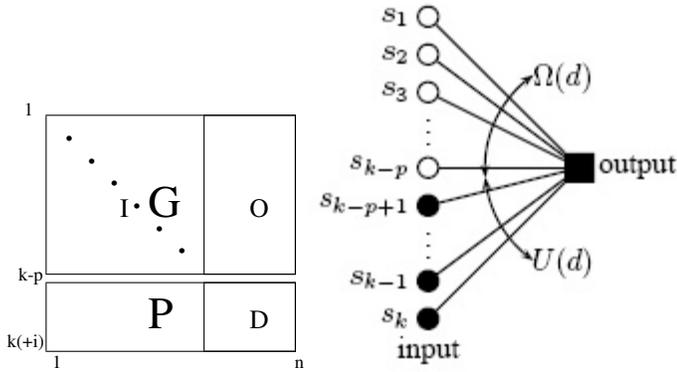
\section{Cost Analysis of Doping with Inactivations}\label{sec:discuss}
\vspace{-1mm}
We consider two performance measures: communication overhead in terms of the percentage of dopings (or their absolute number), and decoding complexity. We treat the overhead in the upfront delivered set of symbols as a parameter, since we assume that the broadcast session duration is determined by design, considering broadcast channel statistics (see our use case example in the next section). The quality of the erasure channel to a particular client is random, and so is the required doping overhead. Given that the upfront delivery was not sufficient, we may decide to dope more or less, depending on the strategy of the cost tradeoff, and how sophisticated our decoding method is. The graphs in Figures~\ref{fig:dopPer}~and~\ref{fig:compersym} depict the tradeoff between overhead and complexity. For a source block of length $k,$ the complexity $C$ is calculated as
\begin{eqnarray}\eqnlabel{costLinGE}
C=\frac{C_l+C_g}{k-d}=\frac{k-p+d +(p + i +u - d)^g}{k-d},
\end{eqnarray}
where $p$ is the number of PIs, $i$ is the number of DIs, $u$ the number of uncovered symbols, and $d$ is the number of dopings we request, $u \leq d\leq u+i$, while $2.5\leq g\leq 3$ is the exponent in the complexity of Gaussian elimination  $C_g(x)=O(x^g)$. The complexity cost is normalized per non-doped source symbol (Fig.~\ref{fig:compersym}).

\begin{figure}[!t] 
\begin{center}
\hspace{-4.5mm}\includegraphics[width=3.2in]{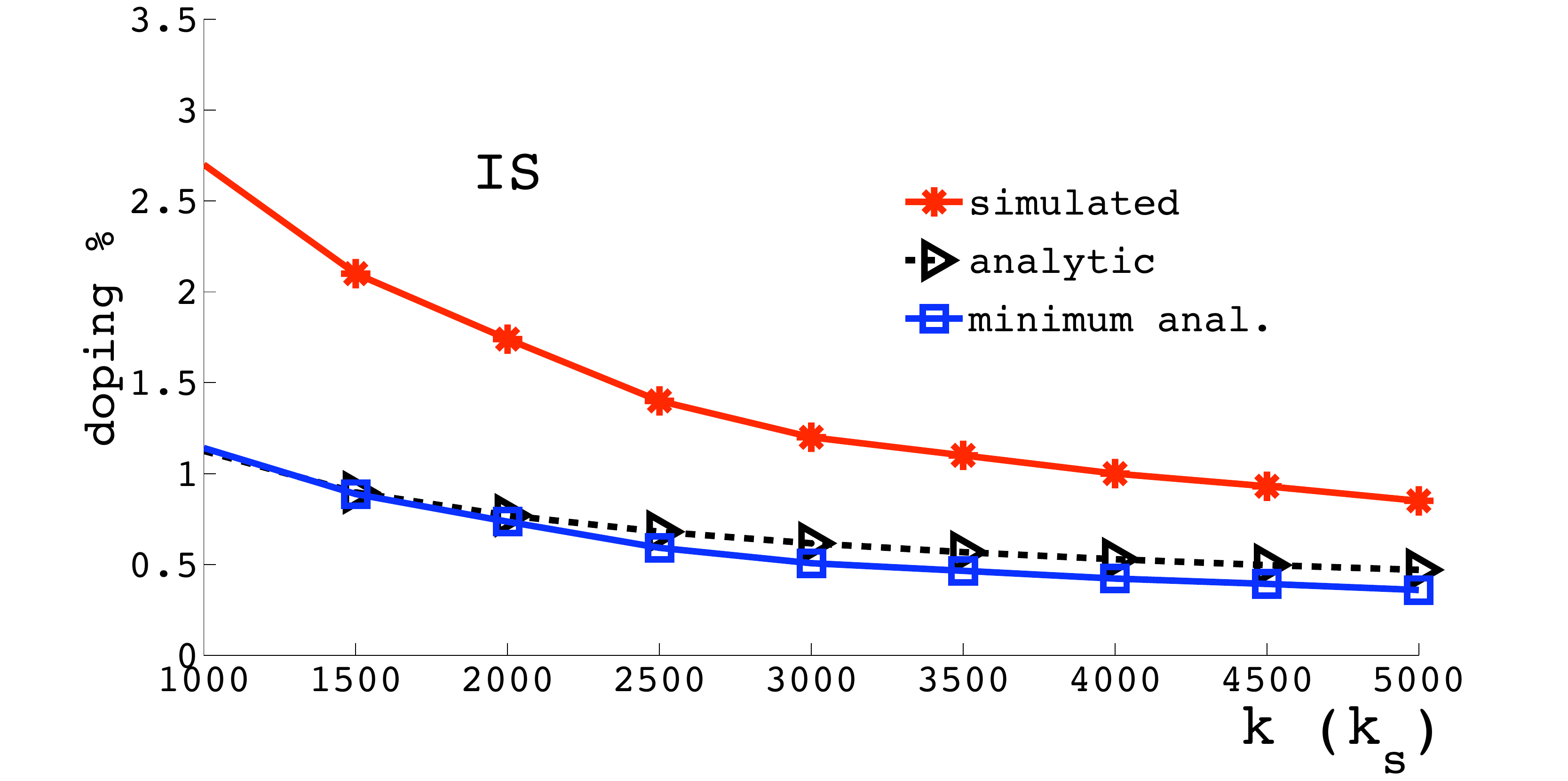}
\end{center}
\vspace{-4mm}
\caption{Doping percentage: the gap to our analytical model (black w/ triangles) is becoming insignificant for $k>5000.$ The blue curve denotes expected minimum number of dopings, based on the simulation curve (red), and \eqnref{frankprob} applied to $D_i.$  Here, $k=k_s$ for all $k.$
}\vspace{-4mm}
\label{fig:syman}
\end{figure}

Note that an estimate of the complexity may be obtained based on the analytical values of the above variables. The lower bound on $i$ is given by the equations \eqnref{Recurs}, \eqnref{parsum}. Evaluation of \eqnref{parsum} for $k=1000$ results in $i \approx 10,$ while for $k=5000,$ $i \approx 222,$ which corresponds to $1\%$ and $0.5\%$ of $k$ (Fig.~\ref{fig:syman}). As $k$ grows, the bound for $i$ is becoming tighter, and its value insignificant (Figure~\ref{fig:syman}). Since the model assumes $k>>\ell,$ the bound is expected to be looser for smaller $k,$ as the finite decoding stages where $\ell\approx k$ have more impact. It is interesting to observe in Figure~\ref{fig:dopPer} that, while for $k=1000,$ the simulation value of $i=3\%$ does not match the bound of $1\%,$ as soon as $k_s=1.05k,$ the number of inactivations hits the $1\%.$ This suggests a possible way of quantifying the impact of the finite $k$ in our model, although this problem is outside the scope of this paper.

The estimate of $u$ is obtained based on the following reasoning.
The probability that a source symbols is not a neighbor of an output node is $1-d_a/k$, where $d_a$ is the average degree of an output node. The probability that a source node is not a neighbor of any output node is  $(1-d_a/k)^{k_s}.$ As the average degree of the output nodes (counting only within the upper subcolumns -- with degrees sampled from IS) is $\log{k-p} \approx \log{k},$ the probability of uncovered nodes is
$(1-\log{k}/k)^ {k(1+\delta)} \approx \eX{-(1+\delta)\log{k}},$ hence,
\begin{eqnarray}\eqnlabel{uncIS}
u \approx k\eX{-(1+\delta)\log{k}}.
\end{eqnarray}
Note that for $k_s=k,$ $u\approx 1$ (see the close-up in Figure~\ref{fig:dopPer}, Section~\ref{sec:discuss}).

For Raptor LT (R-LT), the average degree is some constant, independent of $k.$ For the standardized Raptor distribution \cite{r10}, simulated here, $d_a\approx 4.5.$ Even for $k=1000$, this is significantly lower than  $\log{k},$ resulting in $u \approx k\eX{-4.5(1+\delta)},$ which for $k_s=k,$ saturates $u$ to approximately $0.01k.$ For medium to large $k$ (which is our range of interest), this is significantly larger than for IS.

GE is considered to be of cubic complexity, although there exist methods which leverage the matrix structure, which can slightly lower the exponent. In our graphs we take the lower bound of $2.5$ for the exponent $g$ (which is not tight).
Figure~\ref{fig:compersym} illustrates that for $p\neq 0$ the non-linear complexity term $C_g$ has visible but still moderate effect on the overall complexity, even if the number of dopings $d$ is equal to $i+u$ (red curves with square markers), when this term contributes with the value of $O(k^{1.25}).$
\subsection{Rank Deficiency After Inactivations}\label{subsec:rankdef}
In plain terms, minimum required number $d$ of repair symbols corresponds to the number of equations missing for the upper submatrix to be of full rank $k-p.$ This number is always smaller or equal $u+i,$ as our decoder, for the sake of decoding linearity, inactivates some of the symbols that could be solved by GE.

For $p\neq 0,$ the upper decoding submatrix is a thick matrix even for $k_s=k,$ as it contains $k-p$ rows. When it is of full rank, the number of dopings $d$ may be decreased down to $u,$ if we decide to solve the $i$ inactivated variables through Gaussian elimination. Certainly, for $k_s=1000,$ the slight complexity increase in such cases (min dopings curves, in black, Figure~\ref{fig:compersym}) is due not only to higher $i$ inside the term  $i +u - d$ in the base of $C_g,$ but also due to minimal $d.$

\begin{figure}[!t] 
\begin{center}
\vspace{-6mm}
\hspace{-5.5mm}\includegraphics[width=3.7in]{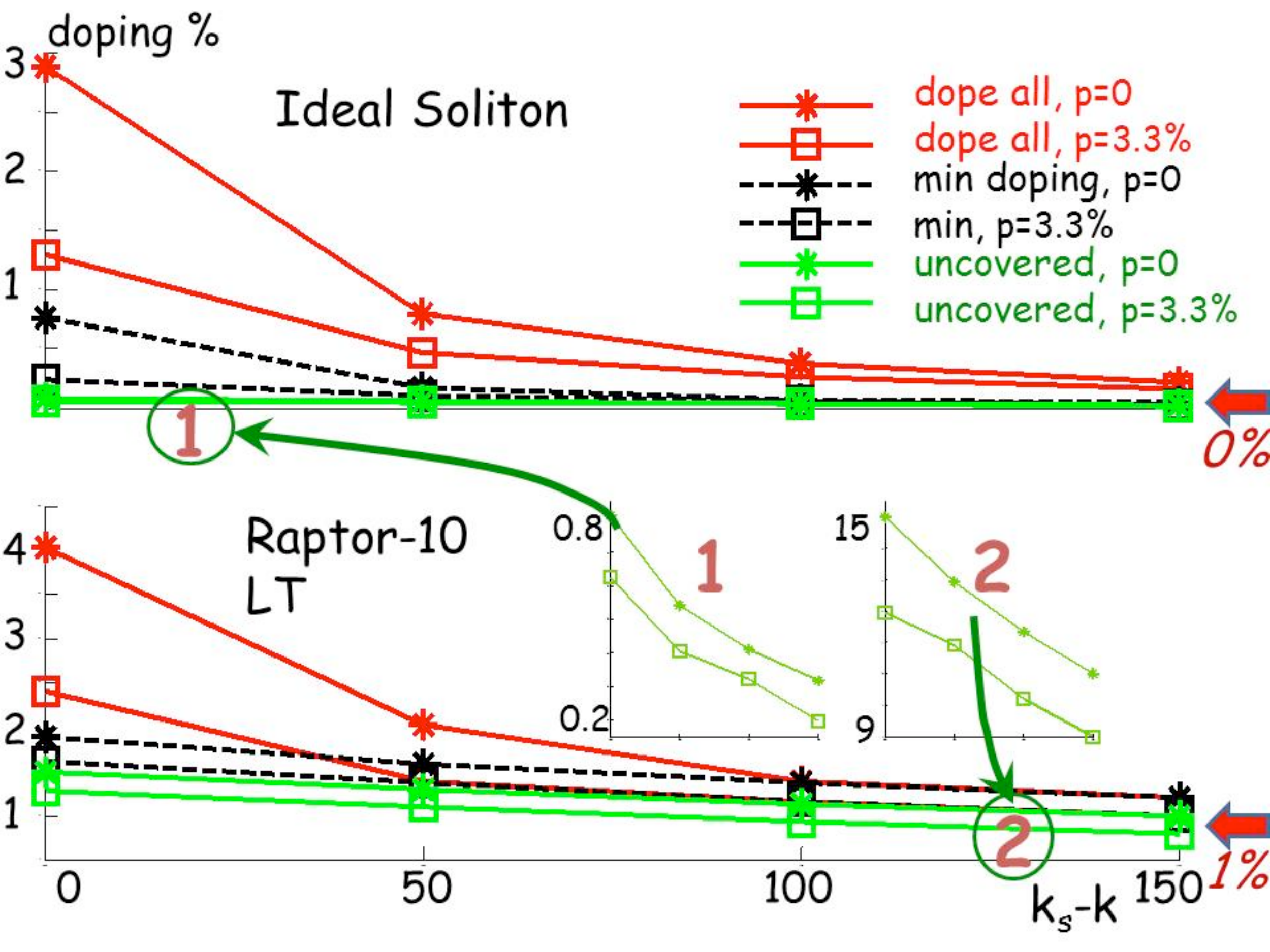}
\end{center}
\caption{Percentage of dopings vs. overhead in collected symbols $k_s-k$ for $k=1000$: the PD in its conditional mode produces smaller repair overhead for the IS based Fountain (upper graph) wrt Raptor-10 based Fountain (lower). The green curves, which represent the percentage of uncovered (not present in any eqn) input symbols that MUST be doped, are magnified in the close-up expressing the absolute values (number of symbols); they illustrate that for Raptor the large number of uncovered symbols does not improve sufficiently with increasing $k_s$ and, hence, saturate the doping percentage above $1\%.$
}
\label{fig:dopPer}
\end{figure}

To estimate the ranks of submatrices involved in decoding, we apply the results from \cite{Cooper2000} and \cite{Kolchin-RG}. They state that there exists a threshold $p(k)$ on the probability of the unit value of an IID (Independent Identically Distributed) binary matrix element, above which the rank sufficiency/defficiency of such a random matrix resembles a completely uniform binary matrix. Consider such a random matrix of size $k\times k+m.$ Let $\alpha=(k+m)/k$  be a constant, $0 < \alpha < \infty.$ Suppose further that $x(k)$ is a function decreasing to $0$ sufficiently slowly with $k.$
Then this probability threshold can be expressed as
\begin{eqnarray}\eqnlabel{pk}
p(k) =\frac{\log{k} + x(k)}{k}.
\end{eqnarray}
Practically, this rank similarity with the purely random matrix holds
provided $p(k)$ does not tend to either zero or one too rapidly. Note now that, for IS-based codes, the average degree of the upper subcolumns is approximately $\log{k}.$ Hence, the number of unit elements is $k_s\log{k},$ and, hence, the probability of ones, under the IID assumption, is $\frac{k_s\log{k}}{k_s(k-p)}.$ For larger $k$, this is sufficiently above the threshold \eqnref{pk} to ensure good rank properties of the upper submatrix. This expectation is confirmed by our simulations, as presented in Figure~~\ref{fig:dopPer}, where the minimum number of dopings for the IS reaches zero for $k_s=1100,$ regardless of the permanent inactivations.
\begin{figure}[!t] 
\begin{center}
\hspace{-4.5mm}\includegraphics[width=3.5in]{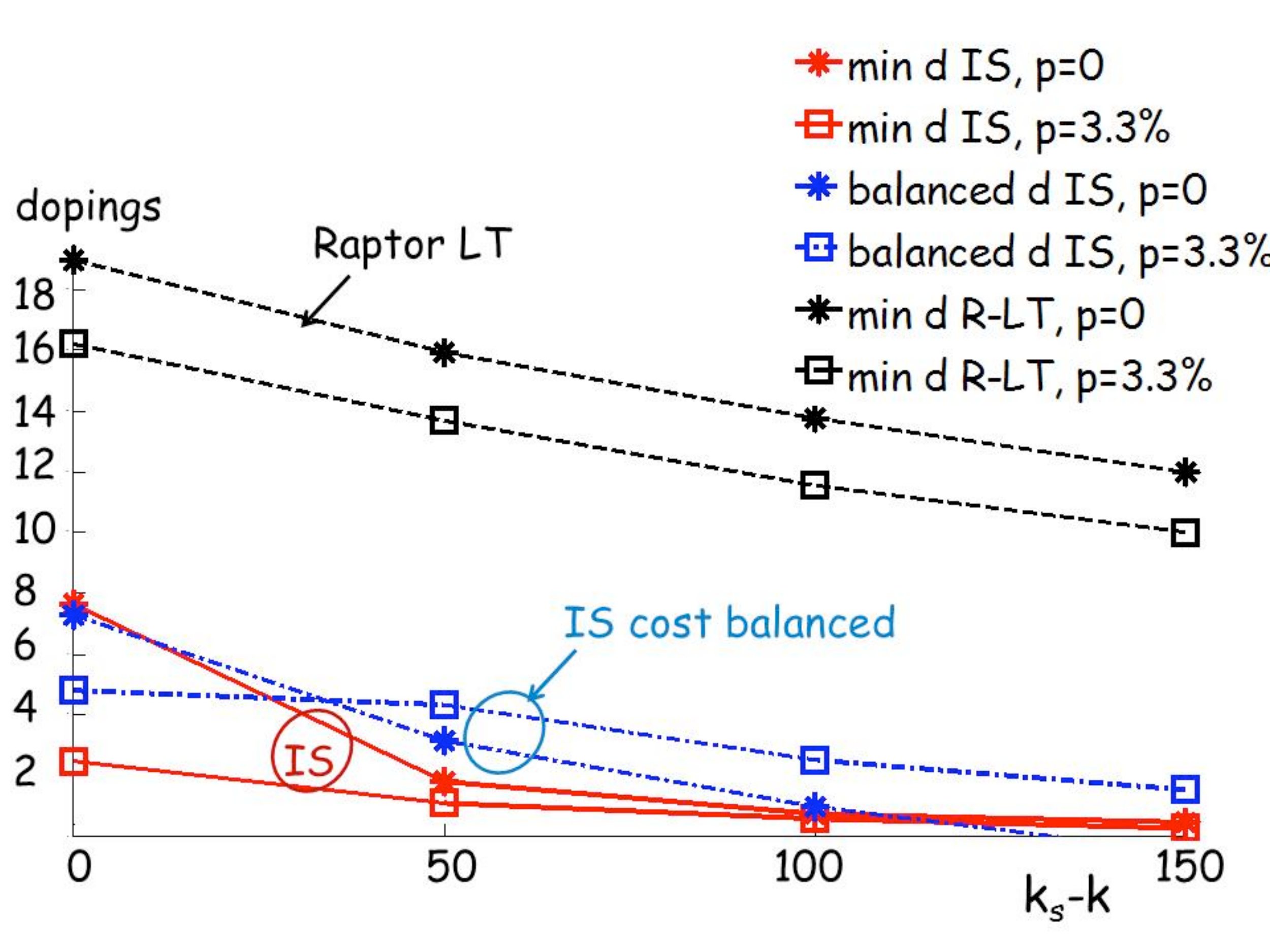}
\end{center}
\vspace{-4mm}
\caption{Minimal number of dopings vs. the overhead in collected symbols $k_s-k,$ for $k=1000$, plotted in {\em red} and {\em black-dashed}, for {\em IS} and {\em Raptor LT}, respectively, against the number of dopings needed for the IS to have {\em equal decoding complexity as the R-LT,} plotted in {\em blue.} The trade-off in the number of additional dopings is not big wrt the IS minimal doping, and dopings are still much below the miniml R-LT amount.
}\vspace{-4mm}
\label{fig:dopPer1}
\end{figure}

\begin{figure}[!t] 
\begin{center}
\hspace{-7mm}\includegraphics[width=3.8in]{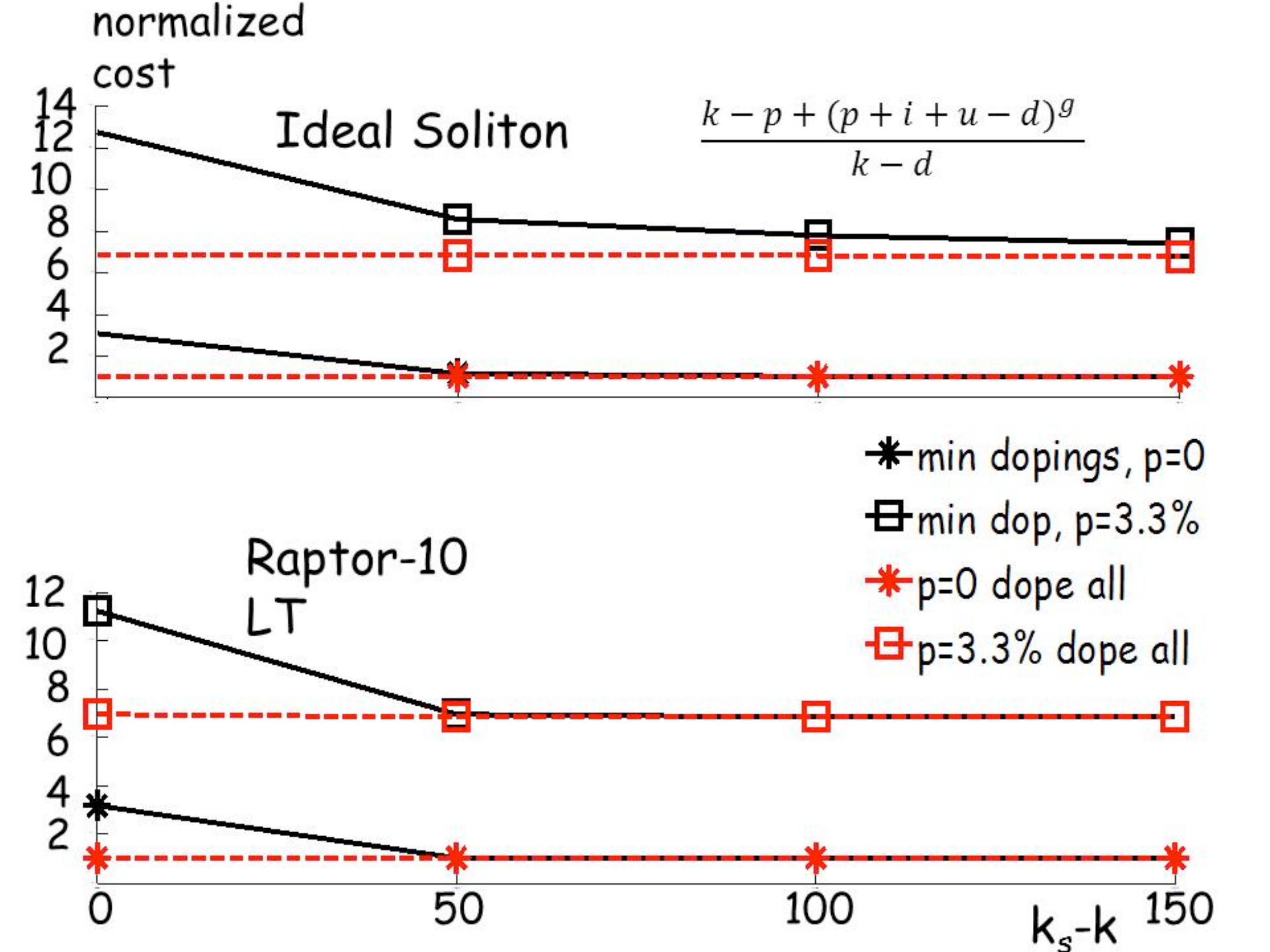}
\vspace{-8mm}
\end{center}
\caption{Complexity per source symbol vs. the overhead in collected symbols $k_s-k$: note that permanent inactivations do bear a price in complexity, as those equations are always solved via GE, but the complexity is still on the same order of magnitude as with iterative decoding only.}
\label{fig:compersym}\vspace{-2mm}
\end{figure}

Differently from the IS-based LT codes, the degree distribution for Raptor-based codes does not depend on $k.$ As $k$ grows, the the density of unit elements $\frac{4.5}{k}$ diverges from the threshold \eqnref{pk}. From the graph perspective, the number of edges in the decoding graph is increasingly insufficient to cover all source symbols, as demonstated by the estimate of $u$ in the previous subsection. In the Raptor design, this relaxation is expected to be compensated for by the pre-code. In the design that insists on simple codes and low decoding delay as here, the uncovered symbols must be recovered through doping. As a baseline, $d=i+u,$ except when the rank of the submatrix $D$ is $p+i,$ when $d=u.$ The importance of the matrix density in terms of the uncovered (must-dope) symbols is illustrated in Figure~\ref{fig:dopPer}. Note that the red curves in Figure~\ref{fig:dopPer} denote doping-only approach which dopes $u$ uncovered and $i$ inactivated symbols, while dashed black curves denote doping of only those symbols that cannot be decoded through GE, hence, the minimal doping. The close-up in the same figure illustrates the inferiority of R-LT codes in terms of $u$, and the doping percentage graphs in the same figure reflect this in the total doping overhead. With Raptor, the uncovered symbols form the  majority in the doping structure, and that is why it has a larger overhead despite the fact that the number of DIs is slightly smaller (for small $\delta$ only). This is because the peeling decoder is applied to $k-p-u$ source symbols only, and $u$ is signifficant. PI has similar effect on both designs - it practically eliminates the occurance of uncovered nodes among last $p$ input symbols. However, even for $p>0,$ the problem of uncovered symbols with R-LT in the upper submatrix becomes more pronounced for larger $k$.

The manner in which ``partial'' GE is performed after the PD has finished has much bearing both on the complexity and on the doping-induced (repair) communication cost. Our simulations with IS show that for $k=1000,$ when $k_s=k(1+\delta)$ and  $\delta \geq 10\%,$ the matrix $D$ is singular in fewer than $0.1\%$ of cases, and for $p=\sqrt(k)$ and $2/3\sqrt(k),$ it happens slightly sooner. Hopefully, if the GE decoding of submatrix D succeeds, both permanently and dynamically inactivated symbols are known without any doping, and $d=u,$ which is vanishing for the IS. 

Let us take a closer look at the reasons for such a high likelihood that the matrix D is of full rank.
The two submatrices of matrix $D,$ $D_p$ and $D_i,$ have different structures.  As mentioned, $D_p$ is a random matrix, formed by the permuted lower subcolumns whose degree is sampled from the uniform distribution $U^p(.).$ Given that additionaly this is is a thick matrix, it is non-singular with very high probability \eqnref{frankprob}. The submatrix $D_i$ is formed by the propagation of left-hand-side (LHS) graph edges belonging to the LHS node connected to the inactivated source symbol (see Figure~\ref{fig:dynact}). The LHS nodes in the IS graph have Poisson degree distribution of mean $\frac{n\log{k}}{k},$ which remains stationary throughout decoding, as the right-hand-side (RHS) nodes maintain the IS distribution. Hence, the average number of unit coefficients propagated to $i$ rows of the submatrix $D_i$ is $i\frac{n\log{k}}{k}.$ Under the IID assumption, we may apply the same reasoning as for the upper submatrix, which is that $D_i$ is non-singular with high probability, based on the threshold \eqnref{pk}. The expected density of $D_i$ is confirmed by the simulations. In addition, Figure~\ref{fig:syman} shows that the number of dopings when $k_s=k,$ and $p=0,$ decreased by the result of \eqnref{frankprob} applied to $D_i,$ matches both the lower bound and the simulations.
\subsection{Performance Comparison: IS vs. Raptor LT}\label{subsec:perfcomp}
\begin{figure}[!t] 
\begin{center}
\hspace{-3.5mm}\includegraphics[width=3.5in]{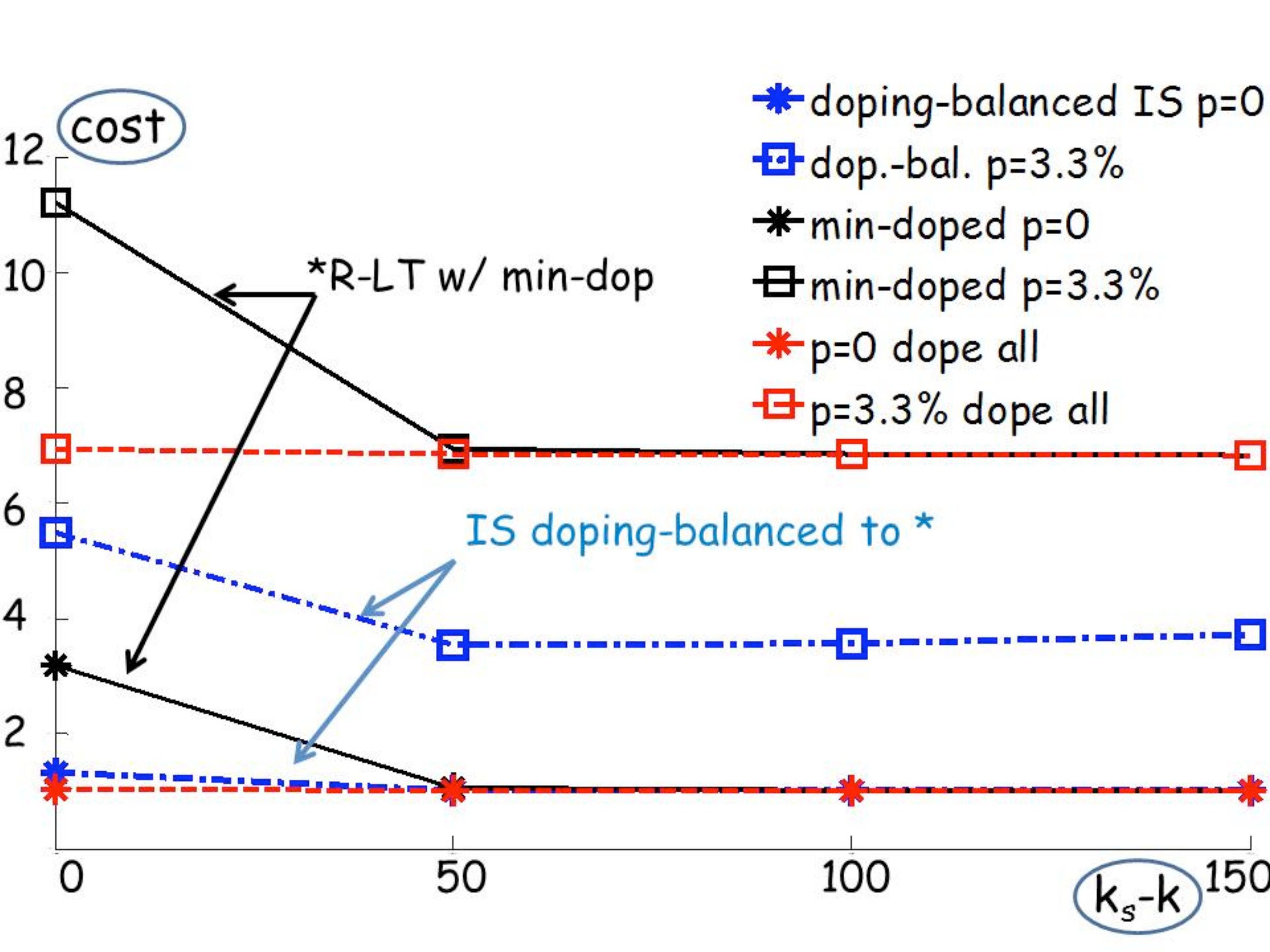}
\end{center}
\vspace{-3mm}
\caption{Cost for the R-LT min and max (dope all) number of dopings vs. the overhead in collected symbols $k_s-k,$ for $k=1000$, plotted in black-dashed and red, respectively, against the cost needed for the IS to have {\em the number of dopings equal to R-LT min,} here plotted in blue.
}
\vspace{-4mm}
\label{fig:compersym1}
\end{figure}

While Figure~\ref{fig:dopPer} clearly illustrates the advantage of the IS based codes in terms of the repair communication cost, Figure~\ref{fig:compersym} may leave the reader under impression that this advantage is taken away by the increased complexity cost with respect to R-LT (graph for minimal doping, when $p=3.3\%$). To illustrate that this is not the case, we introduce Figures~\ref{fig:dopPer1}~and~\ref{fig:compersym1}, which provide a fair performance comparison between the two code variants.
Figure~\ref{fig:dopPer1} plots the minimal doping curves from both subgraphs in Figure~\ref{fig:dopPer} (marked as Raptor LT and IS), against the minimal dopings allowed for the IS variant to achieve exactly the same critical complexity as the Raptor-LT variant (black curves in the lower subgraph of Figure~\ref{fig:compersym}). Marked as {\em IS cost balanced}, these curves show that, when the complexity is the same, the IS doping is still significantly below the minimal R-LT amount, and not much higher that the IS minimum value.
To further illustrate the IS advantage, Figure~\ref{fig:compersym1} plots the R-LT complexity cost (from the lower subgraph of Figure~\ref{fig:compersym}) against the IS complexity cost when the minimum number of dopings is matched to R-LT level (i.e. by the black curves in the lower subgraph of Figure~\ref{fig:dopPer}). Hence, these curves, marked as the {\em IS doping-balanced} curves, are to be compared with the black curves, marked by {\em R-LT w/ min-dop}, where the IS demonstates better performance again.

For good channels (larger $k_s,$ e.g 1150) with IS AL-FEC, the DIs ensure linearity of decoding of an already solvable system of equations, as the value of must-dope symbols $d=u=0$ whp (see the red pointer in the upper subgraph of  Figure~\ref{fig:dopPer}). This is not the case for the R-LT distribution, as indicated by the red pointer in the corner of the lower subgraph of Figure~\ref{fig:dopPer}.

In conclusion, apart from the tractability of their analytical model, the IS based codes are superior in terms of both repair costs, communication and complexity, and their simplicity and true ratelessness may be of further utility with multimedia broadcast scenarios where cooperative peer-to-peer schemes are allowed.
\section{Example Use Case}\label{sec:exam}
Let us assume that the density of multimedia subscribers is about $15\%,$ which is the current addressable market for mobile TV services. As the density of mobile subscribers in urban areas is typically around 300 users, this results in a moderate estimate of about 50 concurrent multimedia clients. Note that these figures are bound to grow, as the current trends are exponential. If the mobile provider desides not to allocate extra bandwidth to AL-FEC, the repair overhead for each user will be equal to the experienced erasure rate. Let us assume that the expected average rate is $\epsilon_d=5\%,$ which is realistic according to \cite{StandardAL-FEC}. This incurs the total repair overhead per cell of $250\%.$ For $k=1000,$ the repair overhead amounts to $2500$ packets, incurring high maximum repair delay.

Now, assume that the proposed approach is used, and the delivery phase duration is designed so that for an average user (i.e. experiencing erasure rate $\epsilon_d$) $k_s\approx 1100.$ This is equivalent to broadcasting $n\approx 1150$ symbols, meaning that the provider accepts the AL-FEC communication overhead of $15\%.$ However, according to \eqnref{parsum},\eqnref{edop}, and confirmed by simulation results presented in Figure~\ref{fig:dopPer}, for $k_s=1100,$ the repair overhead is almost zero, with slightly increased but still linear complexity, and certainly lower than $0.5\%$ with unit per-symbol complexity. Hence, upper bounding the per-user repair overhead to $0.5\%,$ we obtain the total repair overhead per cell of $25\%.$ Hence, the total application layer communication overhead (AL-FEC plus repair) amounts to $40\%,$ as opposed to $250\%$ without AL-FEC.

Given the obvious savings on the provider side, let us consider the effect on a particular priority subscriber to multimedia broadcast streaming. Assume that the application layer buffer stores the next block (a video segment) while the current one is being repaired, and the last one is played out. This allows for a repair time of about half second.
Now, if the erasure rate is as estimated, the repair time will be due to the transfer of at most a couple of packets. If the channel is better, no repair delay will be incurred. For $\epsilon > \epsilon_d,$ e.g. $\epsilon=10\%$ due to mobility and extreme interference, $k_s$ would be approximately $1050,$ which incurs a delay of at most 10 packet transfers. Hence, the buffer size would be sufficient to guarantee steady video quality, without artifacts (i.e. freezing) due to dropped segments. The proposed application of delayed dopings will minimize feedback to one-time request of priority dependent delay $\Delta_f.$
Besides the fact that high priority users would be repaired first, with no waiting time, it is likely that the unicast bearer would offer a better physical channel (PHY) FEC, minimizing the loss of repair packets. Note that the bandwidth overhead due to a stronger PHY FEC is negligible given such a low repair overhead.

Hence, under the assumptions of this use case, the proposed AL-FEC, based on the low-complexity two-phase decoder of a simple IS code, provides both good quality of experience and scalability to large number of multimedia subscribers.
\bibliographystyle{IEEEtran}
\bibliography{isit12}
\end{document}